\documentclass[journal]{IEEEtran}
\usepackage{caption}
\usepackage{graphicx} 
\usepackage{color}
\usepackage{multirow}
\usepackage{tabularx}
\usepackage{amsmath}
\usepackage{amsfonts}
\usepackage{amssymb}
\usepackage{float}
\usepackage{xspace}
\usepackage{pifont}
\usepackage{subcaption}
\usepackage{enumerate}
\usepackage{enumitem}
\begin{document}
\title {Physical-AI: From Channel Awareness to Environmental Intelligence in 6G Wireless Networks}

\author{Farooque Hassan Kumbhar, \textit{Senior Member, IEEE}, Kapal Dev, \textit{Senior Member, IEEE}, Sunder Ali Khowaja, \textit{Senior Member, IEEE}, Alexandros-Apostolos A. Boulogeorgos, \textit{Senior Member, IEEE}, 
Mehdi Bennis, \textit{Fellow, IEEE}, Yuanwei Liu, \textit{Fellow, IEEE}

\thanks{F. H. Kumbhar is with the Augmented Cognition Meta-communications ERC Research Center, Korea University, Korea  (e-mail: farooque.kumbhar@ieee.org)}
\thanks{K. Dev (corresponding author) is with Department of Computer Science, Munster Technological University (MTU), Ireland  (e-mail: kapal.dev@ieee.org)}
\thanks{S. A. Khowaja is with the School of Computing, Faculty of Engineering and Computing, Dublin City University, Ireland (e-mail: 
sunderali.khowaja@dcu.ie)}
\thanks{A. Boulogeorgos is with the Department of Electrical and Computer Engineering of the Democritus University, Greece  (e-mail: al.boulogeorgos@ieee.org)}
\thanks{Mehdi Bennis is with the Centre for Wireless Communications, University of Oulu, 90014 Oulu, Finland (e-mail: mehdi.bennis@oulu.fi)}
\thanks{Yuanwei Liu is with the Department of Electrical and Computer Engineering, The University of Hong Kong, Hong Kong, and also with the Department
of Electronic Engineering, Kyung Hee University, Yongin-si, Gyeonggi-do
17104, South Korea (e-mail: yuanwei@hku.hk).}

\thanks{``The authors would like to acknowledge and extend their deepest regards for the insightful feedback of Prof. Dusit Niyato, Nanyang Technological University, Singapore; regarding the proposed foundation-based model for latent representation.''}
}
\maketitle


\begin{abstract}
Conventional wireless networks rely on instantaneous channel state information (CSI) and react to channel variations without explicitly modeling the physical environment, limiting their ability to handle blockage, mobility, and interference in dynamic deployments. Paradigms such as \emph{Integrated Sensing and Communication (ISAC)} add sensing capabilities but lack explicit environment modeling and decision-making. In this article, we propose Physical-AI: a \emph{new architecture for environment-aware wireless networking}, where radio signals enable sensing, modeling, and interaction with the environment in addition to data transmission. The framework proposes a self-supervised spatiotemporal radio foundation model for transforming distributed radio observations into a shared latent environmental representation. Multiple inference heads operate on this representation to estimate key environmental properties, including blockage, user distribution, mobility dynamics, and interference structure. A task-specific neural decision layer maps this representation to proactive, context-aware control actions. By integrating perception, world modeling, and decision-making in a closed loop, the proposed framework goes beyond ISAC and establishes Physical-AI as a promising architecture for intelligent 6G systems. Simulation results show that the proposed predictive framework reduces outage probability and blockage-response latency, particularly under increasing beam-switching delays.

\end{abstract}

\begin{IEEEkeywords}
Physical-AI, ambient radio sensing, Wi-Fi sensing, channel state information, integrated sensing and communication, 6G, multipath propagation, embodied intelligence.
\end{IEEEkeywords}

\IEEEpeerreviewmaketitle
\section{Introduction}
\label{sec_intro}
Sixth Generation (6G) cellular networks are expected to support futuristic applications and a broader range of heterogeneous device types, including ground and aerial devices with distinct mobility patterns and different sizes~\cite{zhang_general_2023}. It is noteworthy that wireless communication issues remain largely the same, ranging from resource allocation, interference management, and beamforming to sensing and communication. However, the requirements for latency, data rate, pervasive communications, and accessibility have become increasingly stringent. Despite this, modern wireless systems largely treat the propagation medium as a stochastic channel and rely on instantaneous channel state information (CSI) to optimize communication performance~\cite{zhang_general_2023}. The existing paradigm has been improved through data-driven optimizations and the artificial intelligence (AI)-based solutions, but it fundamentally remains channel-centric~\cite{zheng_2026jepamsac}. CSI-enabled decisions only provide a reactive view of the environment and object presence, such as adaptive beamforming, resource scheduling, and power control ~\cite{li_recent_2026, santra_machine_2025}. To support the stringent requirements of modern 6G wireless networks, it is essential to move beyond reactive adaptation toward \emph{environment-aware intelligence} that captures the physical factors shaping radio propagation.

Current wireless networks already support dense and heterogeneous radio deployments, where the environment is populated by numerous transmitters, sensing-capable devices, and distributed infrastructure. The continuous emission of radio signals interacts with devices, users, and obstacles, becoming an information-rich embedding. These signals contain environmental information within the observed wireless channel~\cite{zheng_2026jepamsac, lei_near_field_2025}. Recent advances in integrated sensing and communication (ISAC) have demonstrated that wireless signals can be used to infer properties of the target environment~\cite{santra_machine_2025}. We go beyond signal-inferences and use passive sensing to utilize these embedded, information-rich radio signals to effectively understand environmental dynamics. This well-represented environment-aware information can provide additional context and details to address various wireless communication challenges. We propose a paradigm shift that combines integral concepts from three major aspects, i.e., environment-aware Physical-AI, ISAC, and indoor localization and sensing. 
 
\textbf{Foundations of Physical-AI:} AI for wireless communications has been extensively studied for tasks such as beamforming, resource allocation, interference management, and mobility prediction~\cite{zheng_2026jepamsac, li_recent_2026, santra_machine_2025}. These approaches leverage data-driven models to improve network performance, but operate directly on communication measurements and thus remain largely limited to link-level optimization. However, concepts that go beyond reactive AI-based solutions envision embodied and physical AI in cyber physical systems~\cite{ray_physical_2025}. Moreover, large language models (LLMs) can be used to enhance collective decision-making in a ``perception-action loop"~\cite{feng_multi_agent_2025}. 

\textbf{Integrated sensing and communication (ISAC):} ISAC represents a paradigm shift for 6G, where sensing and communication functionalities share the same spectrum and hardware infrastructure to improve efficiency~\cite{zhang_intelligent_2025, zhang_general_2023}. Existing ISAC research primarily focuses on waveform design and signal processing techniques for estimating environmental parameters such as range, velocity, and direction~\cite{zhang_general_2023}. Successful environment and target sensing have also been shown to improve multiple-input, multiple-output (MIMO) communications, localization~\cite{lei_near_field_2025}, and interference mitigation~\cite{xu_interference_2024}. Transformer-based LLM~\cite{li_recent_2026}, LLMs~\cite{wang_generative_2025}, and Generative AI (GAI) ~\cite{yang_generative_2024} have been incorporated with the ISAC models to address physical layer issues including channel estimation, secure mobile crowd-sensing, beamforming, and resource allocation.

\textbf{Indoor sensing localization and positioning:} There has been extensive AI-based work on exploiting variations in CSI or received signal strength to infer indoor environmental information~\cite{santra_machine_2025}. Indoor localization and positioning information enable applications such as activity recognition, object detection, mobility analysis, and device-free localization~\cite{zhu_intelligent_2026}. Indoor sensing methods use state-of-the-art machine learning (ML) techniques, including physics-informed neural networks (PINNs) for high-fidelity electromagnetic field approximation~\cite{zhang_physics_informed_2025}, meta-learning and graph neural networks to generalize across dynamic scenarios~\cite{xiao_meta_simgnn_2026}, deep supervised dictionary learning for robust device-free localization~\cite{huang_wi_fi_2025}, and few-shot learning for reliable activity classification. While recent machine learning techniques have improved sensing accuracy, these methods are typically task-specific and do not provide a generalizable representation of the environment. On the other hand, there exist several public datasets that include contextual sensing, CSI, received signals, and other information. These datasets include Wi-Fi data from SiMWiSense 2023, Bluetooth data for empty parking spaces from 2022, ultra-wideband signal features from 2022, and CSI and radar data heatmaps from MMVR 2024 and HUPR 2023.

Despite these advances, there exists a critical research gap and a fundamental focus error in how research pursues these wireless communication challenges. Current systems are largely reactive, relying on instantaneous or few-shot CSI and received signal information. This reactive approach ignores the fact that wireless signals inherently encode the structure and dynamics of the physical environment. Most existing models fail to capture the intrinsic physical information embedded within the received signal properties, leading to poor generalization across the environment and its factors, such as user density, blockage, mobility, and interference. This limitation motivates us to rethink the entire solution space by shifting the question from ``what are the channel parameters?" to ``What is the environmental state and cause of channel parameters?".

We propose Physical-AI framework that unifies sensing and communications through a shared latent representation of the physical environment. Our proposed self-supervised spatiotemporal radio foundation model can learns transferable latent environmental states from distributed wireless observations. The proposed framework can capture the physical causes, rather than the symptoms, of channel variations such as blockage, activity, and interference. The shared latent environmental state can be processed using task-specific inference heads to estimate environmental dynamics and generate actionable network control decisions. We design a three-layer architecture that integrates distributed radio sensing, latent environmental modeling through the proposed radio foundation model, and proactive network control.

The main contributions are summarized as follows:
 
\begin{itemize}

\item A new three-layer Physical-AI framework is introduced for wireless systems to infer latent environmental states from distributed radio observations and enable environment-aware communication decisions.

\item We propose a self-supervised spatiotemporal radio foundation model-based architecture that can learn transferable latent environmental states from large-scale wireless observations and supports multiple downstream inference and control tasks.

\item A closed-loop Physical-AI system is presented that integrates environmental inference with proactive network control for beamforming, scheduling, and mobility management.

\item We demonstrate the feasibility of the proposed framework through a predictive environment-aware beam switching scenario.
\end{itemize}

\begin{figure*}[t]
    \centering
    \includegraphics[width=\textwidth]{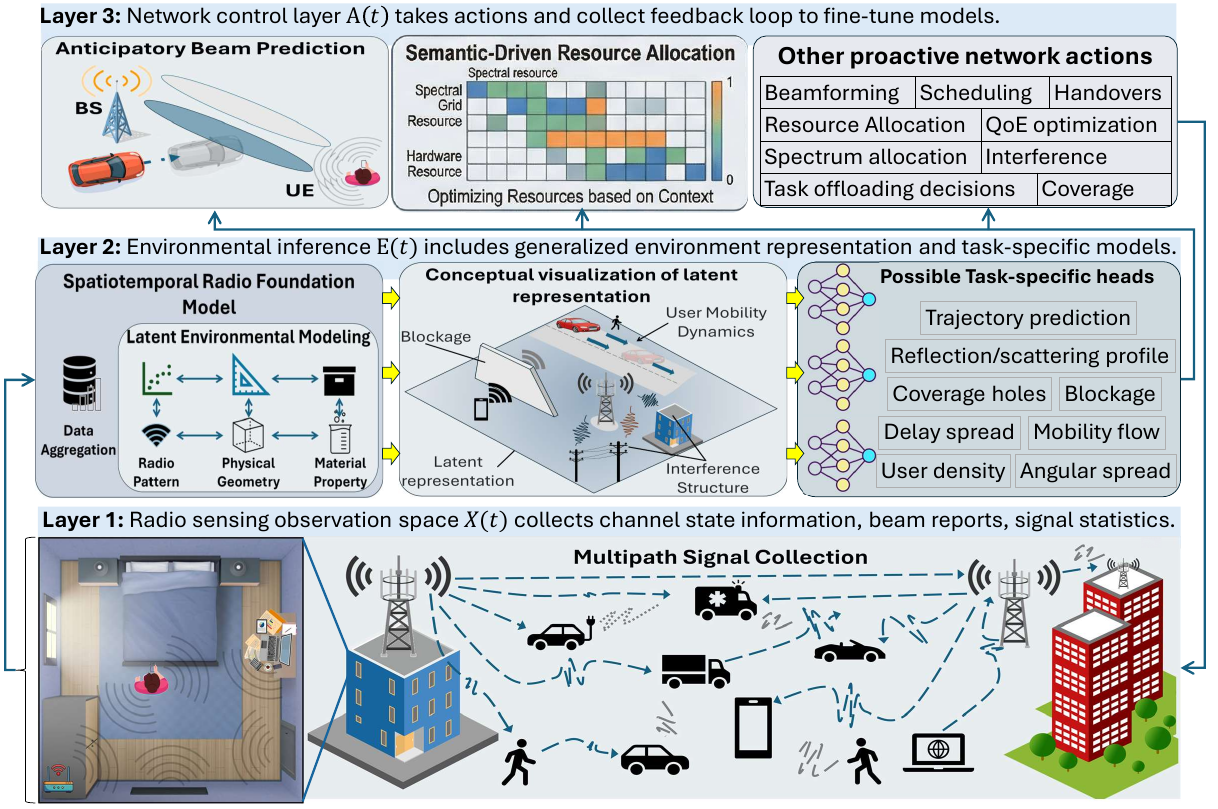}
	\caption{The proposed 3-layer Physical-AI framework of environment-aware wireless networking. }
	\label{fig_environment_aware}
\end{figure*}

\section{System Model and Environmental State Representation}
\label{sec_systemModel}  
We consider a radio-rich wireless environment populated by multiple transmitters, including Wi-Fi access points, cellular base station (BS), and other wireless devices, which continuously emit radio signals. These signals propagate through the environment and interact with surrounding objects, users, and obstacles, inherently embedding information about the physical environment into the observed wireless channel. A sensing-enabled network entity, such as a BS, edge server, or centralized controller, collects radio observations from multiple distributed nodes across the network. These observations provide a multi-node and multi-modal view of the environment, capturing both spatial and temporal variations in radio propagation. Unlike conventional systems that focus solely on communication, the objective here is to leverage these observations to infer environmental states that influence network behavior.

\subsection{Radio Observation Space and Environmental State Representation}
Let ${X}(t)$ denote the collection of radio observations at time $t$. This observation space aggregates measurements from multiple transmitters, receivers, and frequency bands. The collected data may include CSI, received signal strength (RSSI), beam measurements, and temporal channel variations. In practice, these observations are affected by noise, hardware impairments, and synchronization errors. Nevertheless, when aggregated across nodes and time, they provide a rich and redundant representation of the underlying environment. 
The wireless channel is fundamentally governed by the physical environment, including user distributions, blockage conditions, mobility patterns, and interference interactions. To explicitly capture these factors, we define an environmental state that captures the dominant physical factors affecting radio propagation. This state includes components such as spatial user density, obstacle and blockage distribution, mobility flows, and interference topology. Unlike instantaneous channel measurements, the environmental state evolves on a slower time scale and provides a more temporally consistent representation of the system dynamics. This enables the network to anticipate environmental changes rather than only reacting to instantaneous channel degradation. While radio environment maps and channel knowledge maps primarily characterize spatial propagation properties, the proposed environmental state additionally captures temporal environmental dynamics relevant to proactive network control.

\subsection{Limitations of Reactive Wireless Systems}
Conventional wireless systems are designed to directly optimize network control actions, such as beam management and resource scheduling, based on instantaneous channel measurements \cite{li_recent_2026}. The approach is effective for short-term adaptation, but it remains inherently reactive because the wireless channel is treated as a stochastic black box rather than a materially instantiated byproduct of the physical world \cite{ray_physical_2025, li_recent_2026}. Existing ISAC frameworks extend communication systems with sensing capabilities, but they primarily focus on instantaneous environment perception and waveform-level sensing tasks without maintaining a persistent environmental representation. This ``environment-blind" paradigm is a significant problem for next-generation 6G networks, particularly in mmWave and Terahertz (THz) bands. Future 6G networks have high susceptibility to sudden blockages resulting in sharp increases in outage probability and response latency, if systems only react after link degradation is already observed \cite{li_recent_2026, xu_interference_2024}. Existing research on the specific influence mechanisms of the physical environment on radio waves is still insufficient, leading to systems that lack a causal environmental representation~\cite{zhang_general_2023, li_recent_2026}. Most of the designed solutions exhibit poor generalization across diverse or dynamic scenarios. An environment-aware wireless networking paradigm is needed, where radio observations are used not only for instantaneous sensing, but also for maintaining a latent environmental representation that supports predictive and proactive network control~\cite{ray_physical_2025}.

\section{Proposed Environment-Aware Wireless Network Architecture}
\label{sec_proposed}

The core of the proposed framework is the shift from treating the wireless channel as task-specific signal information to viewing it as an information-rich byproduct of the physical world. Conventional systems rely on signal parameters (e.g., CSI and RSSI) for decision-making; however, our vision is to learn a latent world model that captures the dynamics, causal laws, and physics of the environment. The unified latent representation for any given observation space (${X}(t)$), aggregates radio observations, multi-node and multi-band signals, temporal measurements, and semantics. Simply put, the proposed radio foundation model captures the physical causes of signal variations, such as blockage, user mobility, and correlated interference. The proposed architecture consists of three functional layers, with the foundation model as an integral core. Fig.~\ref{fig_environment_aware} illustrates our envisioned three-layer Physical-AI framework, which demonstrates an end-to-end loop from understanding latent representation to network control.

\subsubsection{Radio Sensing Layer}
The first architectural layer is responsible for high-resolution data acquisition and the transformation of raw electromagnetic observations into structured inputs. The radio sensing layer aggregates measurements from multiple transmitters and receivers, including CSI, received signal statistics, beam measurements, and temporal channel features. The radio signal acquisition probes the environment without the need for dedicated sensing transmissions and user participation. The collected information may include temporal, amplitude, and phase data of CSI and RSSI, captured using USRP N321 Software-Defined Radios or commodity Intel 5300 NICs. Raw signal data is required to be reshaped into grid-structured CSI images and multi-channels tensors that can capture spatial multipath patterns. Any hardware offset errors can be reduced through denoising using discrete wavelet transforms~\cite{huang_wi_fi_2025}.

\subsubsection{Environmental Inference Layer}
Given structured radio observations, the second architectural layer (Environmental Inference Layer) functions as the environmental world-modeling module.. Its objective is to estimate a compact latent state that explains the physical causes of radio variations, rather than only predicting communication metrics~\cite{zheng_2026jepamsac, yang_generative_2024}. The input to the model is a time-indexed tensor containing multi-link CSI/RSSI, beam measurements, node locations when available, carrier frequency, and temporal context. The encoder maps this input to a latent representation, for a fixed size observation window. This latent state is designed to capture geometry-related features (dominant scatterers and blockage regions), mobility-related features (user trajectories and flow direction) and radio-interaction features (interference coupling and correlated fading).

The design plan for Layer 2 is to train this model in two stages. First, self-supervised pre-training is performed on large-scale unlabeled radio traces using masked CSI reconstruction, channel prediction, and contrastive learning across different nodes, bands, and time instances. Second, the model is adapted using limited labels or weak supervision, such as camera/LiDAR-assisted blockage maps, floor plans, user locations, or network-side key performance indicators (KPIs). The task-specific inference heads are attached to the shared latent state. These heads estimate spatial user density, blockage probability maps, mobility flows, dominant scatterer/obstacle distribution, and interference topology. The backbone can be implemented using a temporal transformer or graph-temporal encoder, where nodes represent access points, users, or sensing links, and edges represent spatial or interference relationships. This makes the layer scalable from a single-cell deployment to a distributed multi-cell setting.

\subsubsection{Network Control Layer}
The third layer is a task-specific neural network model, built on top of layer 2. It is designed as a plug-and-play inference module that consumes the latent environmental representation produced by the Environmental Inference Layer, together with any task-specific inputs. Network control architecture layer learns and takes best control decisions. Layer 3 learns a mapping from the latent state (capturing geometry, mobility, and radio interactions) to optimal network actions. These actions may include beamforming decisions, scheduling policies, mobility management (e.g., handover), and interference coordination. In this sense, the architecture layer 3 can be interpreted as a learned policy model that directly optimizes system performance based on environmental awareness.

Unlike Layer 2, which focuses on world modeling and multi-task inference, Layer 3 is specialized for a particular control objective and can be independently designed, trained, and deployed. This enables a modular “plug-and-play” architecture, where different neural network models can be integrated depending on the target application. The model can be trained using supervised learning (e.g., from expert policies or historical data). The outcomes of these actions are fed back into the system, forming a closed perception–decision–action loop that continuously improves both the latent representation and the control policy.

\subsection{How It Works: End-to-End Operation}
The proposed closed-loop feedback system unifies perception, cognition, and actuation to solve wireless network challenges. Multiple nodes across the network aggregate high-resolution observations into a structured observation space, which acts as a reflection of the physical geometry of the surrounding environment. The radio foundation model aggregates and processes these observations to produce a shared latent representation of the environment. Task-specific inference heads extract structured components from the latent space, such as blockage risk assessments or mobility dynamics. It should be noted that the task-specific models must be compatible with the control layer and independent of each other, while being trained against different tasks and target labels. The network control layer takes optimal (or near-optimal) actions (e.g., adjusting beams, scheduling resources, or initiating handovers), impacting the wireless environment. The closed feedback loop enables continuous learning, which refines the foundation model and inference heads with the success or failure of the taken actions. The proposed framework allows the network to maintain sustained stability and ultra-low latency, effectively moving towards next-generation 6G ecosystems.

\begin{figure}[t]
    \centering
    \includegraphics[width=0.65\columnwidth]{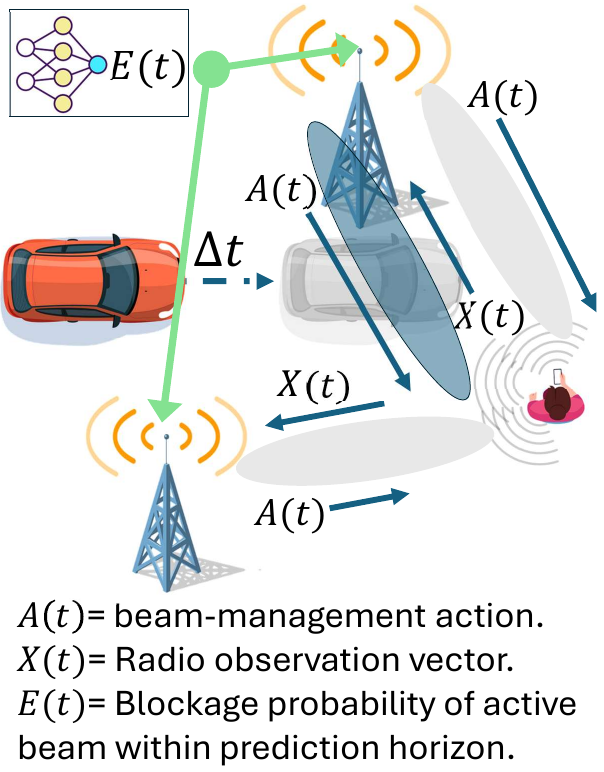}
	\caption{Environment-aware proactive beam-switching.}
	\label{fig_simulation}
\end{figure}

\section{Use Case and Performance Evaluation}
\label{sec_useCase_evalation}
We consider a beam management scenario in a mmWave/THz communication system, where links are sensitive to blockage caused by environmental dynamics, as demonstrated in Fig.~\ref{fig_simulation}. The objective is to maintain reliable connectivity between BS and users by controlling beam-switching or handovers. This use case captures dynamic blockage directly affects link reliability which is a key challenge in mmWave beam management. In a conventional reactive approach, beam switching is triggered only after link degradation is observed through channel measurements. An oracle upper-bound solution has perfect knowledge of the future blockage event, which leads to timely beam-switching action. However, the proposed environment-aware scheme can take predictive beam-switching action based on the estimated blockage risk. Comparing the baseline, predictive, and oracle controllers isolates the benefit of anticipatory beam switching. We chose dynamic blockage because it is a simple beam-management setting in which blocker motion degrades the link before the controller reacts.

The proof-of-concept~\footnote{The implementation code is available upon request by contacting the corresponding author via email.} considers a 28 GHz, 100 MHz BS–UE link over 60 m, with control updates every 10 ms and 6 s episodes. A moving pedestrian blocker crosses the LoS region with randomized crossing time, position, and speed (0.8–2.2 m/s), while $30\%$ of episodes are non-crossing cases to avoid deterministic blockage. Layer 2 receives a 60 ms history window and computes 10 lightweight radio features including current primary and backup SNRs, 10 ms and 30 ms primary-SNR trends, the recent mean/minimum/standard deviation of primary SNR, the recent mean of backup SNR, the instantaneous beam-quality gap, and the recent minimum beam-quality gap. The binary target is whether severe blockage, defined as primary-path attenuation of at least 12 dB, will occur within the next 150 ms. The predictor is trained and tested across multiple randomized episodes, and sweep results are averaged over 10 seeds. We use logistic regression as a lightweight predictor head only for this proof-of-concept, without a base foundation model. The design choice is intentional to show that even stand-alone inference head improve performance. The experiment tests whether short-horizon blockage prediction can improve Layer-3 beam control. For every simulation, blockage trajectories are randomly generated through the crossing position, crossing time, motion direction, and pedestrian speed. At each decision instant, Layer 3 combines the current radio state with the predicted blockage risk to determine whether beam switching should be initiated. The reactive baseline, in contrast, switches only after the primary-beam quality falls below a predefined threshold. 

To evaluate the robustness of the proposed design, four results are considered: (a) outage probability versus switching delay, (b) blockage-response latency versus switching delay, (c) outage probability versus prediction horizon, and (d) outage probability versus risk threshold. The switching-delay analysis captures practical implementation issues in Layer~3, the prediction-horizon analysis reflects the anticipatory capability of Layer~2, and the risk-threshold analysis quantifies the balance between conservative and proactive beam switching. Blockage-response latency is defined as the time elapsed between the onset of severe blockage on the primary link and the instant the controller switches to the backup beam, averaged over all blockage events.

\begin{table}[t]
\caption{Proof-of-concept experimental setup of dynamic blockage in mmWave beam management.}
\label{tab:exp_setup}
\centering
\begin{tabular}{p{3cm}p{4.5cm}}
\hline
Frequency / bandwidth & 28 GHz / 100 MHz \\
Link distance & 60 m \\
Control interval & 10 ms \\
Episode duration & 6 s \\
Training / test episodes & 1000 / 300 \\
Seeds & 10 \\
History window & 60 ms \\
Feature vector & 10 radio features \\
Prediction horizon & 150 ms default, swept from 50--300 ms \\
Predictor & Logistic regression \\
Output & Probability of severe blockage within the next 150 ms \\
Severe blockage definition & Primary-path attenuation $\geq 12$ dB \\
Blocker speed & 0.8--2.2 m/s \\
Non-crossing episodes & 30\% \\
Switch-delay sweep & 0--60 ms \\
Risk-threshold sweep & 0.5--0.9 \\
\hline
\end{tabular}
\end{table}

Figs.~\ref{fig_results}(a) and \ref{fig_results}(b) show the same central effect that the predictive risk estimation converts part of the reaction time into preparation time. As the beam-switching delay increases, the reactive baseline remains exposed to the blocked primary link for longer, which increases both outage probability and blockage-response latency. In contrast, the proposed environment-aware controller initiates switching before severe degradation is fully observed, so it is less sensitive to actuation delay and maintains lower outage and latency. Figs.~\ref{fig_results}(c) and \ref{fig_results}(d) show that the prediction horizon and risk threshold directly shape controller behavior. With a short prediction horizon, the controller has little time to react and its behavior approaches that of the reactive baseline. Increasing the horizon gives Layer 3 more time to prepare for future blockage, although overly long horizons may reduce prediction reliability. The risk threshold controls how aggressively the controller acts on the predicted blockage probability: a lower threshold favors reliability through earlier switching, whereas a higher threshold reduces unnecessary switching at the cost of greater exposure to blockage. Fig.~\ref{fig_results} shows that even a lightweight Layer-2 predictor can improve beam control by giving Layer 3 advance warning of blockage. The prediction horizon and risk threshold act as design controller for balancing early switching against unnecessary delay. The proof-of-concept shows that environment-aware inference becomes valuable when it is coupled to decision making in a closed control loop.

\begin{figure}[t]
	\centering
	\begin{subfigure}{\columnwidth}
		\centering
		\includegraphics[width=0.75\linewidth]{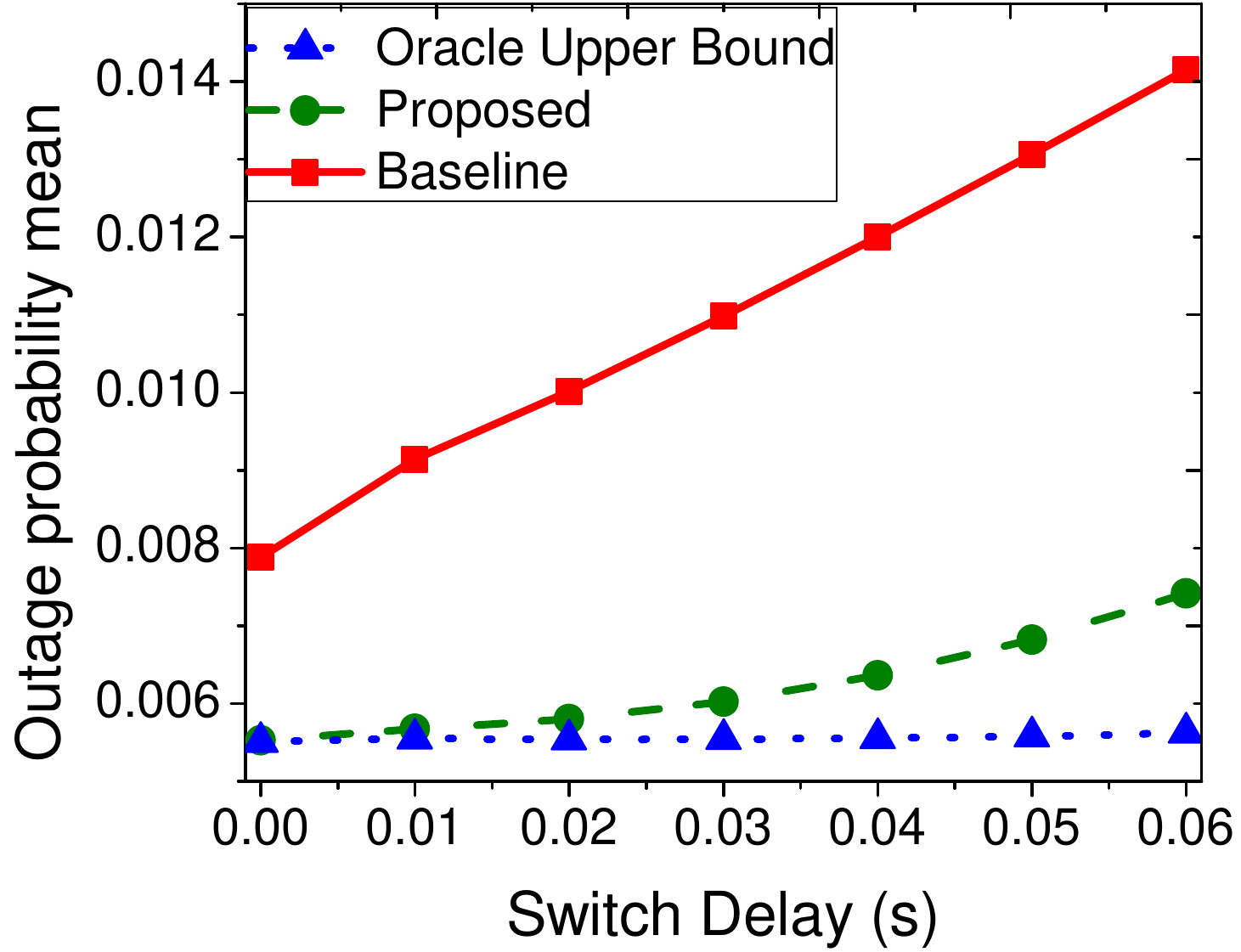}
		\caption{Outage probability versus switching delay}
		\label{fig_switch_delay_outage}
	\end{subfigure}\hfill
	\begin{subfigure}{\columnwidth}
		\centering
		\includegraphics[width=0.75\linewidth]{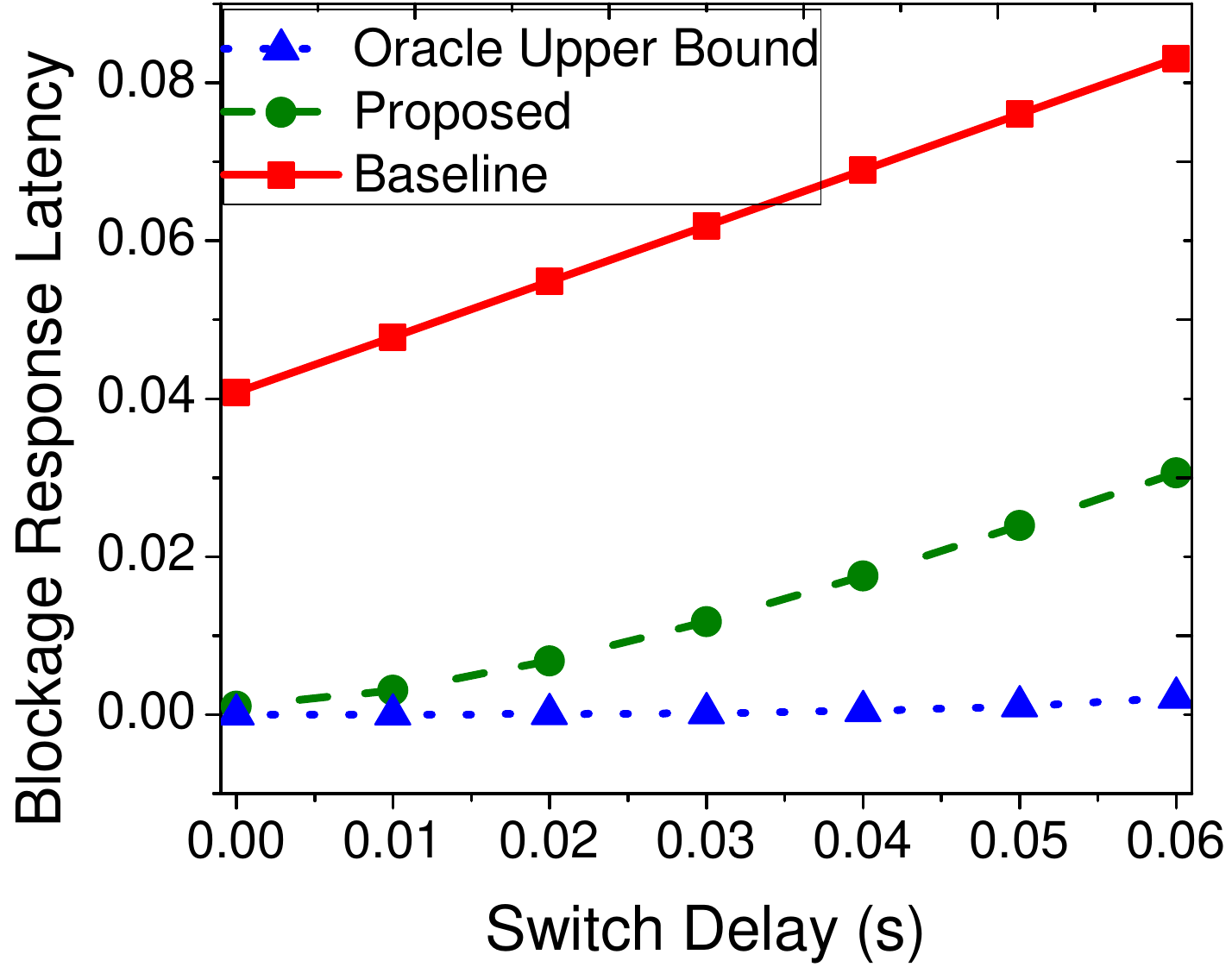}
		\caption{Blockage-response latency versus switching delay }
		\label{fig_switch_delay_latency}
	\end{subfigure}\hfill
	\caption{ Switch delay effect on outage probability and blockage response latency.}
	\label{fig_results}
\end{figure}

\begin{figure}[t]
	\centering
	\begin{subfigure}{\columnwidth}
		\centering
		\includegraphics[width=0.75\linewidth]{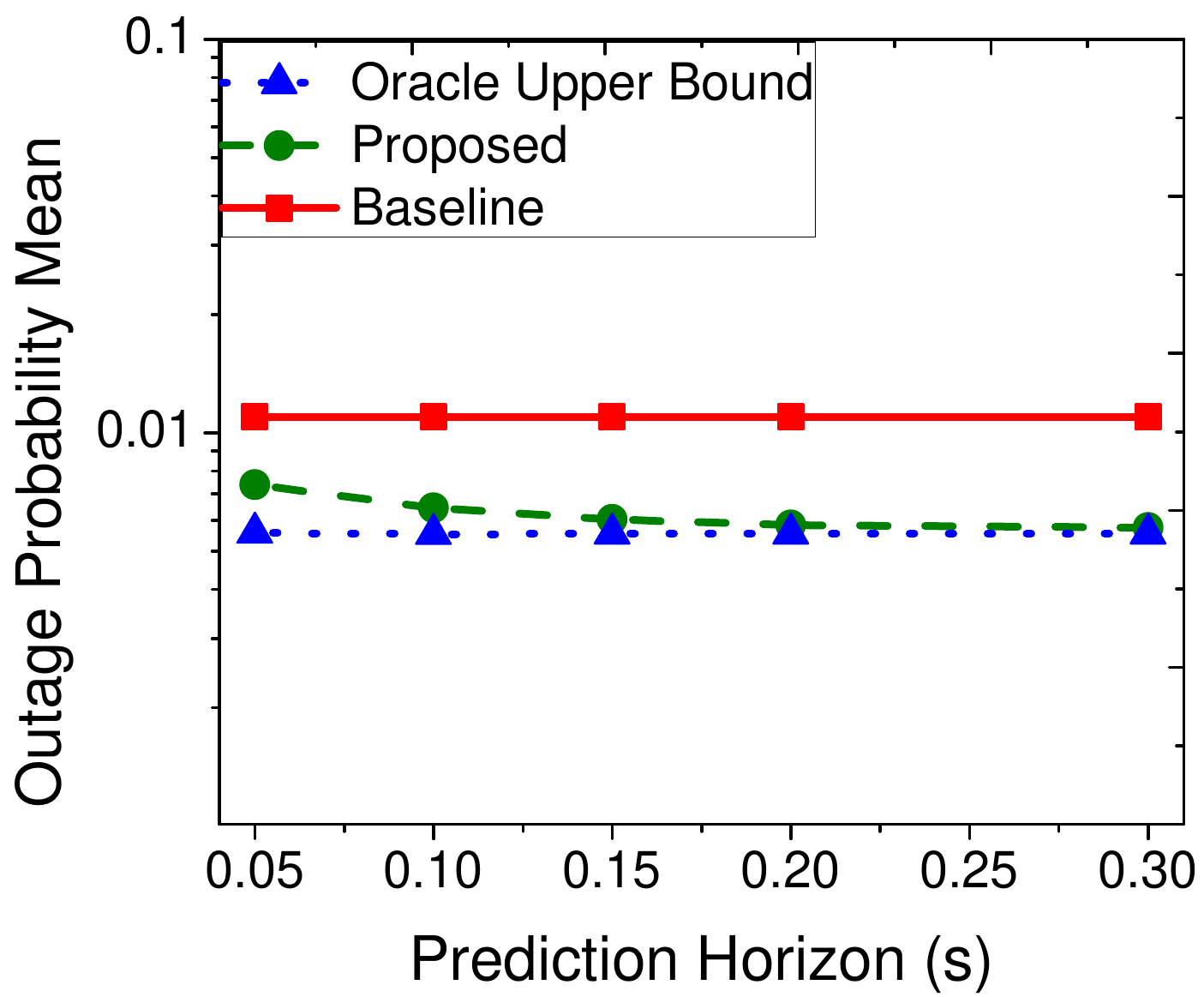}
		\caption{Outage probability versus prediction horizon }
        \label{fig_pred_horizon_outage}
	\end{subfigure}\hfill
    \begin{subfigure}{\columnwidth}
		\centering
		\includegraphics[width=0.75\linewidth]{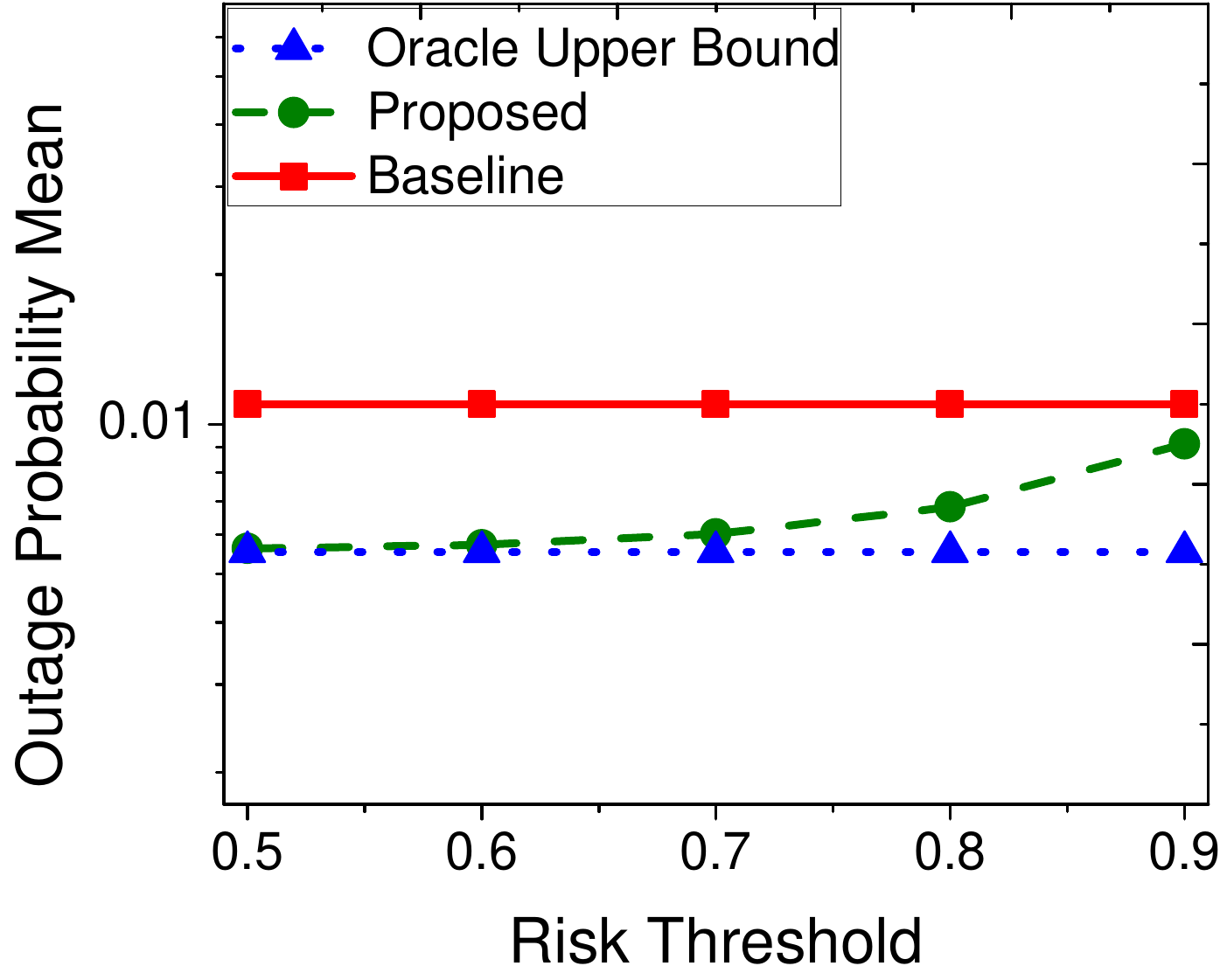}
		\caption{Outage Probability versus risk threshold}
		\label{fig_risk_tradeoff_outage}
	\end{subfigure} 
	\caption{Comparative analysis of beam-switching use case, compared with baseline, proposed, and oracle beam-management schemes.}
	\label{fig_results}
\end{figure}

\section{Open Issues, Challenges, and Future Work}
\label{sec_future}

\subsection{Open Issues}
The proposed Physical-AI framework still raises several unresolved deployment questions. First, it remains unclear how well a shared environment representation can generalize across different sites, frequency bands, node densities, and blockage conditions. Here, the representation is a compact feature space derived from CSI, RSSI, and beam observations. A key open issue is whether one learned representation can support multiple downstream tasks, such as blockage prediction, beam switching, and mobility-aware control, without requiring task-specific redesign for each deployment~\cite{ray_physical_2025, li_recent_2026}. A second issue is how much model complexity is actually needed for radio sensing. Transformer-based models are a reasonable starting point, but their complexity should be supported by measurable gains in inference accuracy, control performance, or latency. Structured attention mechanisms can selectively focus computation on the most relevant antennas, frequencies, or time samples, to improve efficiency and robustness. Similarly, complex-valued neural operators, which process signal amplitude and phase directly, may better reflect propagation physics than purely real-valued models~\cite{zhang_general_2023}. They should be compared against simpler baselines in terms of accuracy, latency, and robustness. Supervision is another unresolved issue, particularly when real labeled measurements are limited. A common practice is that the synthetic and real measurements can be combined, when labeled data are limited~\cite{lei_near_field_2025, ray_physical_2025}. It is especially important for early-stage deployments, where large labeled datasets are unlikely to be available.

\subsection{Challenges}
Practical deployment also faces several engineering challenges. One practical challenge is data collection at scale. RF measurements must cover different indoor and outdoor layouts, mobility patterns, hardware settings, and propagation conditions. Although data augmentation, transfer learning, and synthetic data generation can reduce this burden, the distribution gap between simulated and real environments remains a major concern~\cite{santra_machine_2025}. Joint inference and control under strict latency constraints is another interesting challenge. The framework is only practical if inferred environment information supports timely beam management, handover, and resource allocation under edge latency constraints. In practice, even accurate prediction models may have limited value if inference delay or control overhead is too high for fast channel dynamics~\cite{ray_physical_2025, zhang_intelligent_2025}. An exhaustive study on the system robustness over time is also a possible direction. Wireless environments evolve because of user movement, blockage changes, hardware drift, and site reconfiguration. This motivates continual learning, that is, updating a model over time without retraining it from scratch, and test-time adaptation, where small online adjustments are used to maintain performance during deployment. 

\subsection{Future Work}

Near-term work should prioritize experiments within the current three-layer architecture while accounting for practical deployment constraints. One immediate direction is to develop a shared sensing encoder and evaluate whether partially shared environment representations can support multiple downstream tasks, such as blockage prediction, proactive beam switching, and mobility-aware scheduling, while still allowing lightweight site-specific adaptation~\cite{ray_physical_2025, li_recent_2026}.Another important direction is the development of closed-loop Physical-AI systems in which inferred environment information is continuously fed back to the control layer for real-time decision making. This can be studied using single-agent or multi-agent reinforcement learning, where the learned environment representation serves as a compact state description for beam management, handover, and resource allocation~\cite{feng_multi_agent_2025, li_recent_2026}. These studies should evaluate not only prediction accuracy, but also latency, robustness, and control stability under realistic mobility and blockage conditions. 

Future work should also investigate physics-constrained and uncertainty-aware representations. Embedding propagation consistency, spatial structure, or temporal continuity directly into the learned latent space may improve generalization across frequencies, hardware platforms, and deployment sites. A longer-term direction is to move from task-specific inference toward persistent environment modeling, where the network maintains a continuously updated internal representation of the radio environment over time. Such world-modeling approaches could support predictive beam steering, anticipatory handover, RIS preconfiguration, and mobility-aware resource optimization before channel degradation occurs. Practical studies should further evaluate these frameworks under strict edge execution constraints, including inference latency, energy consumption, online adaptation, and robustness under domain shift~\cite{zhang_intelligent_2025, xu_interference_2024}.
 
\section{Conclusion}
\label{concl}

This article highlights a paradigm shift from \emph{channel-aware} to \emph{environment-aware} wireless networks, where RF sensing and learning-based models enable explicit understanding of the propagation environment rather than relying solely on instantaneous CSI. We proposed a three-layer Physical-AI framework consisting of a radio sensing layer for extracting representations from RF data, an environmental inference layer for deriving semantic knowledge such as blockage and mobility, and a network control layer for translating this knowledge into communication decisions (e.g., handover, beam switching, and resource allocation). The proposed modular design provides a practical pathway to integrate sensing, learning, and control within future 6G wireless systems. By separating large-scale RF data-based representation learning from task-specific optimization, the foundation models enable efficient adaptation across diverse environments and deployment scenarios. We strongly believe that the Physical-AI is expected to become a core framework of AI-native 6G networks, supporting integrated sensing and communication, edge intelligence, and proactive network control.

\bibliographystyle{IEEEtran}
\bibliography{physicalAI_ref.bib}

@article{zhang_general_2023,
	title = {A {General} {Channel} {Model} for {Integrated} {Sensing} and {Communication} {Scenarios}},
	volume = {61},
	copyright = {https://ieeexplore.ieee.org/Xplorehelp/downloads/license-information/IEEE.html},
	issn = {0163-6804, 1558-1896},
	doi = {10.1109/MCOM.001.2200420},
	language = {en},
	number = {5},
	journal = {IEEE Communications Magazine},
	author = {Zhang, Zhengyu and He, Ruisi and Ai, Bo and Yang, Mi and Li, Chao and Mi, Hang and Zhang, Zhangdui},
	month = may,
	year = {2023},
	pages = {68--74},
}

@article{yang_generative_2024,
	title = {Generative {AI} for {Secure} and {Privacy}-{Preserving} {Mobile} {Crowdsensing}},
	volume = {31},
	copyright = {https://ieeexplore.ieee.org/Xplorehelp/downloads/license-information/IEEE.html},
	issn = {1536-1284, 1558-0687},
	doi = {10.1109/MWC.004.2400017},
	language = {en},
	number = {6},
	journal = {IEEE Wireless Communications},
	author = {Yang, Yaoqi and Zhang, Bangning and Guo, Daoxing and Du, Hongyang and Xiong, Zehui and Niyato, Dusit and Han, Zhu},
	month = dec,
	year = {2024},
	pages = {29--38},
}

@article{xu_interference_2024,
	title = {Interference {Mitigation} for {Network}-{Level} {ISAC}: {An} {Optimization} {Perspective}},
	volume = {62},
	copyright = {https://ieeexplore.ieee.org/Xplorehelp/downloads/license-information/IEEE.html},
	issn = {0163-6804, 1558-1896},
	shorttitle = {Interference {Mitigation} for {Network}-{Level} {ISAC}},
	doi = {10.1109/MCOM.001.2300674},
	language = {en},
	number = {9},
	journal = {IEEE Communications Magazine},
	author = {Xu, Dongfang and Xu, Yiming and Zhang, Xin and Yu, Xianghao and Song, Shenghui and Schober, Robert},
	month = sep,
	year = {2024},
	pages = {28--34},
}

@article{lei_near_field_2025,
	title = {Near-{Field} {User} {Localization} and {Channel} {Estimation} for {XL}-{MIMO} {Systems}: {Fundamentals}, {Recent} {Advances}, and {Outlooks}},
	volume = {32},
	copyright = {https://ieeexplore.ieee.org/Xplorehelp/downloads/license-information/IEEE.html},
	issn = {1536-1284, 1558-0687},
	shorttitle = {Near-{Field} {User} {Localization} and {Channel} {Estimation} for {XL}-{MIMO} {Systems}},
	doi = {10.1109/MWC.010.2400237},
	language = {en},
	number = {4},
	journal = {IEEE Wireless Communications},
	author = {Lei, Hao and Zhang, Jiayi and Wang, Zhe and Ai, Bo and Björnson, Emil},
	month = aug,
	year = {2025},
	pages = {190--198},
}

@article{li_recent_2026,
	title = {Recent {Advances} in {Resource} {Allocation} and {Beam} {Prediction} for {Large} {Language} {Models} {Empowered} {ISAC} {Systems}},
	copyright = {https://ieeexplore.ieee.org/Xplorehelp/downloads/license-information/IEEE.html},
	issn = {0163-6804, 1558-1896},
	doi = {10.1109/MCOM.001.2500411},
	language = {en},
	journal = {IEEE Communications Magazine},
	author = {Li, Xingwang and Gao, Yuan and Zeng, Ming and Lei, Xianfu and Hao, Wanming and Nallanathan, Arumugam and Dobre, Octavia A.},
	year = {2026},
	pages = {1--7},
}

@article{wang_generative_2025,
	title = {Generative {AI} {Based} {Secure} {Wireless} {Sensing} for {ISAC} {Networks}},
	volume = {20},
	copyright = {https://ieeexplore.ieee.org/Xplorehelp/downloads/license-information/IEEE.html},
	issn = {1556-6013, 1556-6021},
	doi = {10.1109/TIFS.2025.3570202},
	language = {en},
	journal = {IEEE Transactions on Information Forensics and Security},
	author = {Wang, Jiacheng and Du, Hongyang and Liu, Yinqiu and Sun, Geng and Niyato, Dusit and Mao, Shiwen and In Kim, Dong and Shen, Xuemin},
	year = {2025},
	pages = {5195--5210},
}

@article{zhang_intelligent_2025,
	title = {Intelligent integrated sensing and communication: a survey},
	volume = {68},
	issn = {1674-733X, 1869-1919},
	shorttitle = {Intelligent integrated sensing and communication},
	doi = {10.1007/s11432-024-4205-8},
	language = {en},
	number = {3},
	journal = {Science China Information Sciences},
	author = {Zhang, Jifa and Lu, Weidang and Xing, Chengwen and Zhao, Nan and Al-Dhahir, Naofal and Karagiannidis, George K. and Yang, Xiaoniu},
	month = mar,
	year = {2025},
	pages = {131301},
}

@article{santra_machine_2025,
	title = {Machine {Learning}-{Powered} {Radio} {Frequency} {Sensing}: {A} {Review}},
	volume = {25},
	copyright = {https://ieeexplore.ieee.org/Xplorehelp/downloads/license-information/IEEE.html},
	issn = {1530-437X, 1558-1748, 2379-9153},
	shorttitle = {Machine {Learning}-{Powered} {Radio} {Frequency} {Sensing}},
	doi = {10.1109/JSEN.2025.3547673},
	language = {en},
	number = {13},
	journal = {IEEE Sensors Journal},
	author = {Santra, Avik and Wang, Pu and Shaker, George and Mysore, Bhavani Shankar and Dolmans, Guido and Chen, Yan and Shariati, Negin and Pandharipande, Ashish},
	month = jul,
	year = {2025},
	pages = {23164--23183},
}

@misc{feng_multi_agent_2025,
	title = {Multi-agent {Embodied} {AI}: {Advances} and {Future} {Directions}},
	shorttitle = {Multi-agent {Embodied} {AI}},
	doi = {10.48550/arXiv.2505.05108},
	language = {en},
	publisher = {arXiv},
	author = {Feng, Zhaohan and Xue, Ruiqi and Yuan, Lei and Yu, Yang and Ding, Ning and Liu, Meiqin and Gao, Bingzhao and Sun, Jian and Zheng, Xinhu and Wang, Gang},
	month = jun,
	year = {2025},
	note = {arXiv:2505.05108 [cs]},
}

@article{huang_wi_fi_2025,
	title = {Wi-{Fi} {Sensing} {Based} on {Deep} {Supervised} {Dictionary} {Learning} for {Robust} {Device}-{Free} {Localization}},
	volume = {74},
	copyright = {https://ieeexplore.ieee.org/Xplorehelp/downloads/license-information/IEEE.html},
	issn = {0018-9545, 1939-9359},
	doi = {10.1109/TVT.2025.3552082},
	language = {en},
	number = {8},
	journal = {IEEE Transactions on Vehicular Technology},
	author = {Huang, Huakun and Wang, Chenyang and Zhao, Lingjun and Wang, Weizheng and Ding, Shuxue and Vasilakos, Athanasios},
	month = aug,
	year = {2025},
	pages = {12842--12852},
}

@book{zhu_intelligent_2026,
	series = {{SpringerBriefs} in {Computer} {Science}},
	title = {Intelligent {Localization} for {Integrated} {Sensing} and {Communication}: {Machine} {Learning}-{Driven} {Approaches}},
	copyright = {https://creativecommons.org/licenses/by/4.0},
	isbn = {978-981-96-9384-9 978-981-96-9385-6},
	shorttitle = {Intelligent {Localization} for {Integrated} {Sensing} and {Communication}},
	doi = {10.1007/978-981-96-9385-6},
	language = {en},
	publisher = {Springer Nature Singapore},
	author = {Zhu, Xiaoqiang and Liu, Yuan and Wang, Chunpeng},
	year = {2026},
}

@article{xiao_meta_simgnn_2026,
	title = {Meta-{SimGNN}: {Adaptive} and {Robust} {WiFi} {Localization} {Across} {Dynamic} {Configurations} and {Diverse} {Scenarios}},
	volume = {74},
	copyright = {https://ieeexplore.ieee.org/Xplorehelp/downloads/license-information/IEEE.html},
	issn = {0090-6778, 1558-0857},
	shorttitle = {Meta-{SimGNN}},
	doi = {10.1109/TCOMM.2025.3637041},
	language = {en},
	journal = {IEEE Transactions on Communications},
	author = {Xiao, Qiqi and Ye, Ziqi and He, Yinghui and Liu, Jianwei and Yu, Guanding},
	year = {2026},
	pages = {1732--1746},
}

@article{ray_physical_2025,
	title = {Physical {AI}: {Bridging} the {Sim}-to-{Real} {Divide} {Toward} {Embodied}, {Ethical}, and {Autonomous} {Intelligence}},
    volume={2},  
    DOI={10.1007/s44379-025-00050-y}, 
    number={1}, 
    journal={Machine Learning for Computational Science and Engineering}, 
	author = {Ray, Partha Pratim},
    year={2026}, 
    month={Dec.},
    pages={1–54},
    }

@article{zhang_physics_informed_2025,
	title = {Physics-{Informed} {Neural} {Networks} for {High}-{Fidelity} {Electromagnetic} {Field} {Approximation} in {VLSI} and {RF} {EDA} {Applications}},
    volume={18}, 
    DOI={10.54097/5eqd7y93}, 
    number={2}, 
    journal={Journal of Computing and Electronic Information Management}, 
    author={Zhang, Haijian}, 
    year={2025}, 
    month={Sep.},
    pages={38–46},
    abstractNote={
    The increasing complexity of Very Large Scale Integration (VLSI) circuits and Radio Frequency (RF) systems demands sophisticated electromagnetic field analysis capabilities that traditional Electronic Design Automation (EDA) tools struggle to provide efficiently. This research presents a novel Physics-Informed Neural Network (PINN) framework specifically tailored for high-fidelity electromagnetic field approximation in VLSI and RF Electronic Design Automation applications. The proposed methodology integrates Maxwell’s equations with advanced neural network architectures to enable accurate field prediction across diverse frequency ranges from DC to millimeter-wave operations while maintaining computational efficiency suitable for interactive design workflows. Through comprehensive validation across representative VLSI interconnect structures, RF passive components, and millimeter-wave integrated circuits, our PINN framework demonstrates superior accuracy compared to conventional moment-based methods while achieving computational speedup factors of 45-150× for typical design scenarios. The framework successfully captures complex electromagnetic phenomena including substrate coupling, crosstalk mechanisms, and frequency-dependent losses with mean absolute errors below 3.8% across frequency ranges spanning DC to 300 GHz. The adaptive mesh-free formulation eliminates geometric discretization constraints that limit traditional EDA tools, enabling seamless analysis of irregular conductor geometries and multi-layer dielectric stackups common in advanced semiconductor processes. Real-time field visualization capabilities facilitate intuitive understanding of electromagnetic coupling mechanisms, supporting design optimization workflows that were previously computationally prohibitive. The framework incorporates specialized handling of conductor loss mechanisms, dielectric dispersion effects, and substrate characteristics specific to semiconductor manufacturing processes, ensuring practical relevance for modern VLSI and RF design challenges. Validation against commercial EDA software demonstrates comparable accuracy for standard benchmarks while providing substantial performance advantages for parametric analysis and optimization applications critical to contemporary circuit design methodologies.
    } }

@article{zheng_2026jepamsac,
      title={{JEPA-MSAC: A Joint-Embedding Predictive Architecture for Multimodal Sensing-Assisted Communications}}, 
      author={Can Zheng and Jiguang He and Guofa Cai and Nannan Li and Mehdi Bennis and Henk Wymeersch and Merouane Debbah},
      year={2026},
      volume={2603.29796},
      journal={arXiv},
      primaryClass={eess.SP},
      url={https://arxiv.org/abs/2603.29796}, 
}
\end{document}